\documentclass[preprint,pra,aps,showpacs,preprintnumbers,amsmath,amssymb,eqsecnum]{revtex4}

\begin{document}

\title{Emergent Classicality via Commuting Position and Momentum Operators}

\author{J.J.Halliwell}

\address{Blackett Laboratory, Imperial College,
London SW7 2BZ, UK }


\begin{abstract}
Any account of the emergence of classicality from quantum theory
must address the fact that the quantum operators representing
positions and momenta do not commute, whereas their classical
counterparts suffer no such restrictions. To address this, we
revive an old idea of von Neumann, and seek a pair of commuting
operators $X,P$ which are, in a specific sense, ``close'' to the
canonical non-commuting position and momentum operators, $x,p$.
The construction of such operators is related to the problem of
finding complete sets of orthonormal phase space localized states,
a problem severely limited by the Balian-Low theorem. Here
these limitations are avoided by restricting attention to
situations in which the density matrix is reasonably decohered
(i.e., spread out in phase space).
\end{abstract}

\maketitle

\newcommand\beq{\begin{equation}}
\newcommand\eeq{\end{equation}}
\newcommand\bea{\begin{eqnarray}}
\newcommand\eea{\end{eqnarray}}

\def\z{{\mathbf{z}}}
\def\hP{{ \hat P}}
\def\hp{{ \hat p}}
\def\A{{\cal A}}
\def\D{\Delta}
\def\H{{\cal H}}
\def\E{{\cal E}}
\def\p{\partial}
\def\la{\langle}
\def\ra{\rangle}
\def\ria{\rightarrow}
\def\x{{\bf x}}
\def\y{{\bf y}}
\def\k{{\bf k}}
\def\q{{\bf q}}
\def\p{{\bf p}}
\def\P{{\bf P}}
\def\r{{\bf r}}
\def\s{{\sigma}}
\def\a{\alpha}
\def\b{\beta}
\def\e{\epsilon}
\def\U{\Upsilon}
\def\G{\Gamma}
\def\om{{\omega}}
\def\Tr{{\rm Tr}}
\def\ih{{ \frac {i} { \hbar} }}
\def\trho{{\rho}}

\def\au{{\underline \alpha}}
\def\bu{{\underline \beta}}
\def\pp{{\prime\prime}}
\def\id{{1 \!\! 1 }}
\def\half{\frac {1} {2}}

\def\jjh{j.halliwell@ic.ac.uk}

\section{Introduction}

\subsection{Opening Remarks}

The canonical commutation relation between position and momentum
operators,
\beq
[ \hat x, \hat p ] = i \hbar \label{1.1}
\eeq
lies at the heart of all quantum phenomena.
Physically, this relation corresponds to the fact that
measurements of position and momentum depend on the order of the
measurements. It is this relation, in essence, that is responsible
for the key differences between classical and quantum mechanics,
since in classical mechanics, measurements of position and
momentum can be made in such a way that the order makes no
difference. It follows that any account of the emergence of
classical behaviour from quantum theory must reconcile these two
very different aspects of classical and quantum theory. A typical
point of view is that classical mechanics emerges only at a very
coarse-grained level and for sufficiently coarse-grained samplings
of position and momentum, their non-commutativity makes little
difference \cite{Hal1}.

Still, one wonders whether there is a deeper or more precise way of
reconciling the non-commuting quantum operators with
their commuting classical counterparts. Indeed, this clearly troubled
the founders of quantum theory, since von Neumann addressed the issue in
his 1932 book, Mathematical Foundations of Quantum Mechanics \cite{Von}.
He noted that when we make observations of a macroscopic system,
we are in fact able to make observations of position and momentum
simultaneously (although imprecisely, of course). This suggested to him
that the measurements we make do not in fact correspond directly
to the usual operators $\hat x$ and $\hat p$, but to some other
operators, $\hat X$ and $\hat P$, say, which commute
\beq
[\hat X, \hat P] = 0
\label{1.2}
\eeq
and which must in some sense be ``close'' to the original operators
$\hat x$, $\hat p$. Such a pair of operators could be particularly useful
in bringing a degree of precision to discussions of emergent classicality.
The aim of this paper is to discuss the construction and utility
of such operators.

\subsection{The Imprecision of Classical Physics}

To what degree do macroscropic measurements fix the position and
momentum operators in quantum mechanics? When we make macroscopic
measurements of, say, a particle, there will be imprecisions $
\Delta x $, $ \Delta p $ in the specifications of position and
momentum. These imprecisions may be ``small'' to classical eyes
but they will typically be very large compared to the quantum
scale. It will be useful for what follows to get a quantitative
idea of this. Suppose that the position is measured to a precision
of $10^{-8}m$ and the velocity to within $10^{-8} m/s$ (both extremely
precise specifications, by macroscopic standards). For a mass
of, say,  $10^{-6} kg$, we then have,
\beq
\frac{ \Delta p \Delta x} { \hbar } \ \sim \ 10^{12}
\label{1.3a}
\eeq
This means that when we specify the phase space location of a classical system
in a way that is very precise to classical eyes, at the quantum scale
it could be in any one of about $10^{12}$ phase space cells. There is therefore
a considerable amount of freedom at the quantum level to redefine the position and
momentum operators without making any noticeable difference to macroscopic
observations. Perhaps within this freedom there is the possibility to find
position and momentum operators which commute.

\subsection{Motivations}

What are the motivations for constructing commuting operators
which are close to the position and momentum operators?
An important
indicator of emergent classicality is approximate diagonality of
the density operator $\rho$ (often referred to as decoherence) \cite{JoZ}.
For example, for a point particle
interacting with a thermal environment, it may be shown that after
a short time scale, the density operator approaches the form
\beq
\rho = \int dp dq \ f (p,q) \ | \psi_{pq} \rangle \langle \psi_{pq} |
\eeq
where $ f(p,q)$ is a non-negative function and $ | \psi_{pq} \rangle $
are a set of phase space localized states (such as generalized coherent
states) \cite{HaZ}. This indicates that it is approximately diagonal in both position
or momentum. (It cannot be exactly diagonal in both, since they do not commute.)
This means that there is negligible interference between different values
of position and momenta and, loosely speaking, they may then be treated as
if they are classical. The statement is imprecise, however, since
there is still {\it some} interference, so the variables are only
imprecisely defined.

One would like to be able to make a more exact statement about
diagonality of $\rho$ and this is possible in terms of the commuting operators
$X,P$. For example, using the eigenstates $ | \psi_{nm} \rangle$
of $X,P$ one could construct a density operator of the form
\beq
\rho' = \sum_{nm} f_{nm} | \psi_{nm} \rangle  \langle \psi_{nm} |
\eeq
which is exactly diagonal in $X$ and $P$, indicating there
is exactly {\it zero} interference between different values of these quantities.
One can then choose the coefficients $f_{nm}$ to make
$\rho'$ as close as possible to $\rho$. The question of the degree of
approximate diagonality is therefore shifted to the question of the closeness of the density
operator to a pseudo-classical density operator $\rho'$.

A more comprehensive approach to emergent classicality is the decoherent
histories approach \cite{GH1,GH2,Gri,Omn1,Omn2,Hal2}.
There, the central object of interest is the
decoherence functional,
\begin{equation}
D (\au, \au') = {\rm Tr} \left( P_{\a_n} (t_n) \cdots P_{\a_1}
(t_1) \rho P_{\a_1'} (t_1) \cdots
P_{\a_n'} (t_n) \right)
\label{1.X}
\end{equation}
The histories are characterized by the initial state $ \rho $
and by the strings of projection operators $P_{\a} (t)$
(in the Heisenberg picture) at times
$t_1$ to $t_n$ (and $\au$ denotes the string of alternatives $\a_1
\cdots \a_n$). Intuitively, the decoherence functional is a
measure of the interference between pairs of histories $\au$,
$\au'$. When
\beq
{\rm Re}  D (\au, \au') = 0
\eeq
for $\au \ne \au' $, we say that the
histories are consistent and probabilities  $ p (\au ) = D (\au,
\au ) $ obeying the usual probability sum rules may be assigned to
them. (Typically, the physical mechanisms producing consistency
actually cause the stronger condition  $D (\au, \au') = 0$
for $\au \ne \au'$ to be satisfied, which is referred to as decoherence.)
One can then ask whether these probabilities are strongly
peaked about trajectories obeying classical equations of motion,
and if they are, we can say that the system is emergently classical.

For histories in which the projections $P_{\a_k} (t_k) $
are onto positions at different times of a point particle interacting
with a thermal environment, the decoherence functional, like the density
operator, is {\it approximately} diagonal \cite{GH2,Hal2}. The approximation
is typically exceptionally good, but still only approximate.
Again one wonders whether more exact statements can be made.
Indeed, it has been conjectured that approximately consistent histories
can be in some sense distorted into exactly consistent ones \cite{DoK}.
There are a number of ways in which a set of histories could be
distorted: one can change the initial state, the times of the projections,
the widths of the projections or the operators who spectrum is projected
onto. The possible existence of the commuting variables
$\hat P$, $\hat X$ suggests a particular way of distorting
the histories so as to make them exactly decoherent.

Each position at time $t$ is, in the Heisenberg picture, a function
of the canonical pair, $\hat x, \hat p$, so $\hat x_t = f_t (\hat x, \hat p)$.
The projections onto $\hat x_t$ at different times do not
commute, since $ [ \hat x_t, \hat x_t'] \ne 0  $ in general. The decoherence
functional is therefore not diagonal in general, but can be approximately
diagonal if the system is coupled to a thermal environment.

Now suppose we consider projections onto the variables
$\hat X_t = f_t ( \hat X, \hat P) $ at different times. Under
reasonable dynamics, $\hat X_t$ will be close to $\hat x_t$ as long as $\hat x, \hat p$
are close to $ \hat X, \hat P$. The
operators $\hat X_t $ {\it do} commute at different times so all the projections
commute and, as is easy to see, the decoherence functional will be
exactly diagonal. So exact decoherence is achieved by shifting the operators
$ \hat x, \hat p$ to the commuting pair $\hat X, \hat P$. Furthermore,
it is known that
the probabilities for histories of positions $ \hat x_t$ are typically
peaked about classical evolution. The probabilities for the commuting
variables $\hat X_t$ will therefore have the same property if the $\hat X_t$ are close
to $\hat x_t$. The question of
approximate decoherence and emergent classicality
is therefore shifted to the question of the closeness of the old and new operators.

These then, at least in outline, are the reasons why a commuting pair
of position and momentum-like operators may be useful for discussing
emergent classicality. We turn now to the construction of such operators.

\subsection{Von Neumann's Construction}

Von Neumann outlined a prescription whereby
the commuting operators $ \hat X$ and $ \hat P$
may be constructed \cite{Von}. This involved first taking
a discrete subset $ | m, n \rangle $ of the coherent states $ | p,q \rangle$,
with one state per cell of size $ 2 \pi \hbar $ (a von Neumann lattice).
He alleged that these states are complete (this was later proved
\cite{Per,BBGK,BGZ}).
He then stated that they may be orthogonalized using the Schmidt process to
produce an orthonormal set $ | \psi_{nm} \rangle $. From these,
he constructed position and momentum-like operators
\beq
\hat X = \sum_{nm} n a  | \psi_{nm} \rangle \langle \psi_{nm} |, \ \ \ \
\hat P = \sum_{nm} m \frac {2 \pi \hbar} {a}  | \psi_{nm} \rangle \langle \psi_{nm} |
\label{1.3}
\eeq
(where $a$ is a constant with the dimension of length). These operators
clearly commute. He argued that these new operators are indeed ``close''
to the old ones, in the sense that
\beq
 \langle \psi_{nm} | (\hat x - \hat X)^2 | \psi_{nm} \rangle
 \ \langle \psi_{nm} | (\hat p - \hat P)^2 | \psi_{nm} \rangle
\le  K^2 \hbar^2
\label{1.4}
\eeq
Von Neumann's calculations imply that
the constant $K$ is about $1,800$ (but he thought more detailed calculations
could give a smaller value).

However, von Neumann's prescription is at best at sketch of how
this works and he certainly did not give full details (such as the explicit
form of the $| \psi_{nm} \rangle$).
Furthermore, we now know a lot more about phase space localized
states than was known in 1932, and, as will be described below,
there are obstructions
to constructing such states.
These obstructions do not necessarily apply to what von Neumann did,
but nevertheless, it is still interesting to revisit his ideas from
a more modern perspective.

\subsection{A General Approach and the Balian-Low Theorem}

We start with a more general statement of the problem.
We consider a set of states of the form
\beq
| \psi_{nm} \rangle = U_{nm} | \psi \rangle
\label{1.5}
\eeq
where $ | \psi \rangle $ is a fiducial state and $U_{nm}$
is the unitary shift operator,
\beq
U_{nm} = \exp \left( \ih n a \hat p - \ih m b \hat x \right)
\label{1.6}
\eeq
A particularly interesting case is that of a von Neumann lattice, in which
case $ b = 2 \pi \hbar / a $ and the translations in the $p$ and $x$ directions
then commute, and we have
\beq
U_{nm} = (-1)^{mn} \
\exp \left( \ih n a \hat p \right) \ \exp \left( - i \frac {2 \pi m} {a} \hat x \right)
\label{1.7}
\eeq
There is then one state per cell of size $2 \pi \hbar$, as in von Neumann's case.
It is of interest to find a fiducial state $ | \psi \rangle$ in Eq.(\ref{1.5}) such that
the states are complete and orthonormal, that is,
\bea
\sum_{nm} | \psi_{nm} \rangle \langle \psi_{nm} | &=&  1
\label{1.8}\\
\langle \psi_{nm} | \psi_{n'm'} \rangle &=& \delta_{nn'} \delta_{mm'}
\label{1.9}
\eea
Fiducial states leading to states satisfying these properties are easily
found and we will exhibit a set below. There is, however, a crucial difficulty. According to
the theorem of Balian and Low \cite{Bal,Low,Bat}, if the three properties
(\ref{1.5}), (\ref{1.8}) and (\ref{1.9}) are satisfied,
then the fiducial state $ | \psi \rangle $ has either $ (\Delta x)^2$
or $ (\Delta p)^2$ infinite, so is not phase space localized.
If we used such states to construct commuting operators
$\hat X, \hat P$ as von Neumann did,
then at least one of the averages in Eq.(\ref{1.4})
would diverge, and there would
be no sense in which the new operators are close to the old. (Note that von Neumann
claims to have used the Schmidt procedure to construct his orthonormal set, which
one would not expect to produce states satisfying Eq.(\ref{1.5}), so his construction
does not necessarily fall foul of the Balian-Low theorem).

The problem of constructing orthonormal phase space localized states
is one of great interest in a number of fields so some effort has been expended in finding
ways around the Balian-Low theorem. Zak has proved some interesting results
in this area. In Ref \cite{Zak1}, he showed that the coherent states restricted
to a von Neumann lattice $ | m,n \rangle$ obey a sort of orthogonality relation if the
usual inner product $ \langle m,n | m',n' \rangle $ is averaged over a
single phase space cell. It is not yet clear if this result can be used to produce
commuting position and momentum operators. He has also considered complete
orthonormal sets of states which are localized in position, but double-peaked in momentum
(so localized in $p^2$, but not in $p$. From these one can construct
commuting operators $\hat X$ and $\hat P^2$, which are ``close'' to $\hat x$ and $\hat p^2$
\cite{Zak2}.
This is tantalizing close to the goal
of this paper, but not quite there (and also suggests that the $p \rightarrow -p$ transformation
plays a crucial role in the Balian-Low theorem). Many of these and similar results are proved
using the so-called $kq$ representation, a technique
which is particularly well-adapted to these problems \cite{Zak3}.

To avoid the Balian-Low theorem, one has to drop one of the three
requirements (\ref{1.5}), (\ref{1.8}) and (\ref{1.9}) in order to get
phase space localization.
For the purposes of this paper, which is to construct useful
commuting position and momentum operators, the orthogonality
Eq.(\ref{1.9})
is essential. We will therefore explore the possibility of dropping
the other two requirements. Eq.(\ref{1.5}), the requirement that the states
be obtained by translation of a single fiducial state is mainly for practical
convenience, so there is no harm in relaxing this as long
as the resulting states are not unmanageable.

More significantly,
we will relax the requirement of completeness, Eq.(\ref{1.8}).
The motive behind this is as follows. We are interested in
using commuting operators to discuss emergent classicality. In practice, this
means that we are only concerned with the physical situation in which
the density matrix of the system has
undergone a degree of decoherence. This means that it is approximately
diagonal in both position and momentum, or equivalently, its Wigner function
is reasonably spread out
in phase space. The density matrix is therefore insensitive to the fine
structure of phase space and it seems reasonable to suppose that
the physics could be well-described by a less than complete set of
states, if they are carefully chosen. Of course, this is a quantitative
matter that needs to be checked in detail and we will do this.

For the purposes of constructing commuting position and momentum operators,
we will actually use a construction slightly more general than the one
indicated by von Neumann in Eq.(\ref{1.3}). In particular, we will look
for commuting operators $\hat X$, $\hat P$ of the form
\bea
\hat X &=& \sum_{nm} X_n E_{nm}
\label{1.10}
\\
\hat P &=& \sum_{nm} P_m E_{nm}
\label{1.11}
\eea
Here, $X_n$ and $P_m$ are $c$-numbers and $E_{nm}$ are
projection operators localized onto a region of phase space
labelled by $n,m$ which will consist of {\it more than one}
$ 2 \pi \hbar $-sized cell. They are exclusive
\beq
E_{nm} E_{n'm'} = E_{nm} \delta_{nn'} \delta_{mm'}
\label{1.12}
\eeq
and exhaustive
\beq
\sum_{nm} E_{nm} = 1
\label{1.13}
\eeq
We will also insist that they
are obtained from unitary shifts of  a single projector $E$,
\beq
E_{nm} = U_{nm} E  \left(U_{nm} \right)^{\dag}
\label{1.14}
\eeq
Here, $U_{nm}$ shifts from one cell to the next, so is of the form Eq.(\ref{1.6}),
with $a,b$ chosen so that the translation in the position and momentum directions
commute, but $ ab > 2 \pi \hbar$.
These three conditions are the natural generalization for
projection operators of the requirements (\ref{1.5}), (\ref{1.8})
and (\ref{1.9}), and the original case is obtained with
the choice
\beq
E_{nm} = | \psi_{nm} \rangle \langle \psi_{nm} |
\label{1.15}
\eeq
The quantity $ {\rm Tr} E $ is a measure of the number of $2 \pi \hbar$-sized
cells projected onto, so clearly $ {\rm Tr} E = 1$ in the pure state case,
but $ {\rm Tr} E \gg 1$ in our case, as we will see.

One might have thought that this more general construction could avoid
the Balian-Low theorem, since the restrictions on $E_{nm}$ are in fact
weaker than the restrictions
(\ref{1.5}), (\ref{1.8}) and (\ref{1.9}) for pure states. This is not in
fact the case. We
will prove a modest extension of the Balian-Low theorem which shows
that there is no phase space localized projector $E$ satisfying
the three requirements Eqs.(\ref{1.12}), (\ref{1.13})
and (\ref{1.14}). However, what we will do is find an ``almost'' complete
set of phase space localized pure states from which we can construct
a projector $E$ that satisfies  the exhaustivity condition
Eq.(\ref{1.13}) to a good approximation when acting on sufficiently
decohered density operators. From this we can construct commuting
operators $\hat X$ and $\hat P$ may with useful properties.

\subsection{Earlier Work}

Finally, we briefly mention a related approach. An earlier attempt to
construct commuting position and momentum operators was considered in
Ref.\cite{Hal4}. This construction involved doubling the original Hilbert space
and then using operators defined on this enlarged space. (See also Ref.\cite{HaRo}
for similar ideas). The resulting theory is essentially the same as 't Hooft's
deterministic quantum theory \cite{Hoo}. However, this is no longer standard
quantum theory. In the present work, by contrast, we stay within the framework
of standard quantum theory.

\subsection{This Paper}

In Section 2, we briefly summarize some known properties of the Wigner function
which help to make precise the idea that the state is sufficiently spread
out in phase space. We also briefly note that the Wigner function naturally
suggests an alternative method of defining commuting position and momentum operators.
In Section 3, we prove a modest generalization of the Balian-Low
theorem for projector operators.
In Section 4 we introduce an orthornormal set of phase space localized states
that are ``almost'' complete. We then use them to construct a set of phase
space localized projection
operators $E_{nm}$ which are almost exhaustive.
In Section 5 it is shown that the incompleteness
does not matter if the density operator of the system is reasonably spread out in
phase space. In Section 6, we use the projection operators $E_{nm}$
to construct commuting position
and momentum operators and compute the ``distance'' between these operators
and the usual canonical pair, as in Eq.(\ref{1.4}).
In Section 7, we compare the probabilities for $X$ and $P$ with those for
the usual position and momentum operators and find them to be close.
We summarize and conclude in Section 8. This contribution is based on
the published work Ref.\cite{Hal8}.

\section{Some Properties of the Wigner Function}

It will be useful for the rest of the paper to briefly summarize here some aspects
of quantum mechanics in phase space and the Wigner function. Most of this is standard
material and may be skipped by the informed reader, except the brief comments at the end
of this section.

The Wigner representation for a density operator $\rho$ (or indeed
a wide class of operators) is defined by
\beq
W(p,q) = \frac { 1} { 2 \pi \hbar} \int d \xi \ e^{-\ih p \xi}
\ \rho( q + \half \xi, q - \half \xi)
\label{X2.1}
\eeq
with inverse
\beq
\rho(x,y) = \int dp \ e^{\ih p (x-y) } \ W ( p, \frac {x+y} {2} )
\label{X2.2}
\eeq
Many calculations involving operators are usefully expressed in the Wigner representation.
For example,
\beq
{\rm Tr} \left( A B \right) = 2 \pi \hbar \int dp dq \ W_A (p,q) W_B (p,q)
\label{X2.3}
\eeq
where $W_A (p,q)$ and $W_B (p,q)$ are the Wigner functions of $A$ and $B$.

We will be interested in later sections in the behaviour of the density operator
or Wigner function in a simple model of the decoherence process.
We take the simplest case of a single particle coupled to a thermal
environment in the limit of high temperature and negligible
dissipation, with no external potential. The master equation
for the density matrix $ \rho (x,y) $ is,
\begin{equation}
 \frac {\partial \rho} { \partial t}
= \frac { i \hbar }{2 m} \left( \frac { \partial^2 \rho } {
\partial x^2} - \frac {\partial^2 \rho }{  \partial y^2} \right) -
\frac {D} {\hbar^2} (x-y)^2 \rho
\label{X2.4}
\end{equation}
where $D = 2 m \gamma k T$. In the Wigner representation, the
corresponding Wigner function obeys the equation,
\begin{equation}
\frac {\partial W} {\partial t} = - \frac {p} {m} \frac {\partial W} {\partial x}
+ D \frac {\partial^2 W} {\partial p^2}
\label{X2.5}
\end{equation}

Evolution according to the master equation Eq.(\ref{X2.4}) tends to produce
approximate diagonality in position and momentum. In the Wigner representation,
this appears as a spreading out phase space. Indeed, using Eq.(\ref{X2.5}) it may
be shown that
\begin{eqnarray}
(\Delta p)_t^2 &=& 2 D t + (\Delta p)_0^2
\label{X2.6}\\
(\Delta x)_t^2 &=& \frac {2} {3} \frac {D t^3 } {m^2} + (\Delta p)_0^2 \frac{t^2} {m^2}
+ \frac {2} {m} \sigma (x,p ) + (\Delta x)_0^2
\label{X2.7}
\end{eqnarray}
where
\begin{equation}
\s (x,p) = \half \langle \hat x \hat p +\hat p\hat x \rangle - \langle \hat x \rangle
\langle \hat p \rangle
\label{X2.8}
\end{equation}
evaluated in the initial state. In particular, for long times, the phase space spreading
behaves according to
\beq
\frac{ (\Delta p)_t (\Delta x)_t } { \hbar } \ \sim \ \left( \frac { \gamma k T} { \hbar} \right) t^2
\label{X2.9}
\eeq
This means that any initial state becomes spread out in phase space on the (typically very short)
timescale $ ( \hbar / \gamma k T )^{1/2} $. (See Ref.\cite{AnHa} for similar calculations).

In the case above, a free particle, the spreading continues indefinitely. However, for a bound
system with dissipation, equilibrium is eventually reached. For a simple harmonic oscillator
at thermal equilibrium, for example, the ratio of thermal to quantum fluctuations is
\beq
\frac{ (\Delta p) (\Delta x) } { \hbar } \approx \frac {kT} {\hbar \omega}
\label{X2.10}
\eeq
At typical laboratory temperatures and frequencies that are not unrealistically
fast (for macroscopic systems), this number can be very large, of order $10^{10}$
say. These estimates will be relevant later on. In brief, they show that the
density operator becomes very spread out in phase space (and hence slowly varying)
very readily.

This much is known material and will be used later. However, we now note that the existence of the
Wigner representation suggests an alternative method for constructing commuting position and
momentum operators. It is well-known that when the Wigner function is sufficiently spread out,
it becomes positive and may be regarded as a probability distribution for the variables $p$
and $q$ (which clearly commute, since they are numbers, not operators). The variables $p$
and $q$ do not correspond exactly to the operators $\hat p$ and $\hat x$, but they are
close when the Wigner function is spread out. In fact, it is easy to see that we have the following
correspondences:
\bea
q W (p,q)  \ & \leftrightarrow & \ \frac{1} {2} \left( \hat x \rho + \rho \hat x \right)
\\
p W (p,q) \ & \leftrightarrow & \ \frac{1} {2} \left( \hat p \rho + \rho \hat p \right)
\eea
That is, multiplication by $q$ or $p$ in the Wigner representation corresponds to
the operation of anticommutation with $\hat x$ or $\hat p$ on the density operator
(and it is easy to show that these two operations on $\rho$ commute). The point here is
that these operators on $\rho$ are operations on the space of density operators which
have no counterpart in terms of operations on pure states. This is therefore not
a route to producing the desired pair of commuting operators envisage by von Neumann.

\section{An Extension of the Balian-Low Theorem}

We first show that there is no projection operator satisfying
the three properties Eqs.(\ref{1.13}), (\ref{1.14}) and (\ref{1.15}).
This is an almost trivial extension of the proof of the Balian-Low
theorem given by Battle \cite{Bat}.

We consider the object $ \Tr ( E x p ) $, which we assume exists.
(If it does not, i.e., is infinite, this means that $E$ has infinite
dispersion in either $x$ or $p$, so is not phase space localized).
We have, from the three properties of $E_{mn}$, Eqs.(\ref{1.12})-(\ref{1.14}),
\begin{eqnarray}
 \Tr ( E x p )&=& \sum_{mn} \Tr \left( E x E_{mn} p \right)
\nonumber \\
&=& \sum_{mn} \Tr \left( E x U_{mn} E U_{mn}^{\dag} p \right)
\nonumber \\
&=& \sum_{mn} \Tr \left( E U_{mn} ( x + n a ) E ( p + b m )U_{mn}^{\dag}  \right)
\nonumber \\
&=& \sum_{mn} \Tr \left( E_{-m,-n} ( x + n a ) E ( p + b m ) \right)
\nonumber \\
&=& \sum_{mn} \Tr \left( E_{-m,-n}  x  E  p  \right)
\nonumber \\
&=& \Tr \left( x E p \right)
\nonumber \\
&=& \Tr \left( E p x \right)
\end{eqnarray}
Or in other words,
\beq
\Tr \left( E [ x, p ] \right) = 0
\eeq
which, via the commutation relations, implies that $\Tr E = 0$. Since
$E \ge 0 $, this means that $E = 0$.
So no projector satisfying the three properties exists,
if one insists that they be satisfied exactly. Hence, as indicated
we will relax the conditions in what follows.

We note in passing that this simple result has implications for the
Balian-Low theorem in the pure state case. It may seem that one could
avoid the Balian-Low theorem by dropping the requirement Eq.(\ref{1.5})
and requiring that the states $ |\psi_{nm} \rangle$ are generated
from more than one fiducial state. However, one could then use those fiducial
states (assuming they are orthogonal) to construct a projection
operator $E$ satisfying the properties Eqs.(\ref{1.12})-(\ref{1.14})
and the above result shows that the Balian-Low theorem is not in fact avoided.

\section{An Almost Complete Set of Orthonormal Phase Space Localized States}

We now show how to construct an almost complete orthonormal set of phase space
localized states, from which we can construct the projector $E$.
Our starting point is the set of states considered by Low \cite{Low},
\beq
\psi_{nm} (x) =
a^{-\half} h( x - na) \ e^{2 \pi i m x / a}
\label{2.1}
\eeq
where $h(x)$ a window function on $[-\half a, \half a]$.
These clearly satisfy the three conditions,
Eqs.(\ref{1.5}), (\ref{1.8}) and (\ref{1.9}). The Fourier
transform of these wave function is
\beq
\tilde \psi_{nm} (p) =
\left( \frac{2 a} {\pi \hbar} \right)^{\half} \ (-1)^m \ \frac {
\sin (pa / 2 \hbar) } { (pa / \hbar - 2 \pi m)} \ e ^{- \ih n a p }
\label{2.2}
\eeq
from which it is easy to see that $ (\Delta p)^2$ diverges
because $  \tilde \psi_{nm} (p)  $ goes to zero like $1/p$
for large $p$ which is not fast enough. Differently put, the
window function in Eq.(\ref{2.1}) causes the derivative of the wave function to
involve the $\delta$-functions $ \delta ( x - n a \pm \half a ) $,
which are not square-integrable. These properties are fully
in line with the Balian-Low theorem.

However, inspired by a suggestion of Zak \cite{Zak4},
it is easy to see from Eq.(\ref{2.2}) that we may take linear
combinations of these states which fall off like $1/p^2$ and therefore have
finite $ (\Delta p)^2$. In particular, the states
\beq
\psi^{(1)}_{nm} (x) = \frac {1} {\sqrt{2}} \left( \psi_{n,2m} (x) + \psi_{n,2m+1} (x) \right)
\eeq
have this property. One can also see this in configuration space:
they vanish at $x = na \pm \half a $,
and the offending $\delta$-function appearing in their derivative
therefore causes no problems.
They are not complete, consisting of just ``half'' the states in the momentum
direction, and indeed the states missed out are the states
\beq
\chi^{(1)}_{nm} (x) = \frac {1} {\sqrt{2}} \left( \psi_{n,2m} (x) - \psi_{n,2m+1} (x) \right)
\eeq
and these have infinite dispersion in $p$.

But it is not hard to see that the ``remainder'' states $\chi^{(1)}_{nm}$ decay
like $1/p$ in momentum space, so a further process of ``halving'' is possible
to produce more states with finite dispersion. That is, we define
\beq
\psi^{(2)}_{nm} =  \frac {1} {\sqrt{2}} \left( \chi^{(1)}_{n,2m} (x) - \chi^{(1)}_{n,2m+1} (x) \right)
\eeq
with new remainder states
\beq
\chi^{(2)}_{nm} =  \frac {1} {\sqrt{2}} \left( \chi^{(1)}_{n,2m} (x) + \chi^{(1)}_{n,2m+1} (x) \right)
\eeq
So the set of states $\psi^{(1)}_{nm}$,  $\psi^{(2)}_{nm}$, $\chi^{(2)}_{nm}$ is complete
and orthonormal, but the states $\psi^{(1)}_{nm}$ have infinite dispersion.

We can continue in this way to define
the sequence of states
\beq
\psi^{(K+1)}_{nm} = \frac {1} {\sqrt{2} } \left( \chi^{(K)}_{n,2m+1} - \chi^{(K)}_{n,2m} \right)
\eeq
and
\beq
\chi^{(K+1)}_{nm} = \frac {1} {\sqrt{2} } \left( \chi^{(K)}_{n,2m+1} + \chi^{(K)}_{n,2m} \right)
\eeq
for $K =1,2,3, \cdots $. If we truncate the sequence at some finite value of $K$, $K=N$, say, then the
set of states $ \psi^{(1)}_{nm}, \psi^{(2)}_{nm} \cdots \psi^{(N)}_{nm}$ together with
the remainder states $\chi^{(N)}_{nm}$ are orthonormal and complete.
We therefore have the completeness relation
\beq
\sum_{K=1}^N \sum_{n,m} \ \psi^{(K)}_{nm} (x)  {\psi^{(K)}_{nm}} (y)^*
+  \sum_{n,m}\ \chi^{(N)}_{nm} (x)  {\chi^{(N)}_{nm}} (y)^* = \delta (x-y)
\label{2.10}
\eeq
In some sense, ``most'' of the states, namely the $\psi^{(K)}_{nm} $,
have finite dispersion, and ``some'' of them, namely the $\chi^{(N)}_{nm}$ have infinite
dispersion. In this way, as $N$ increases, the infinite dispersion anticipated from the Balian-Low
theorem is pushed into a progressively smaller set of states. (An interesting question
is whether the limit $N \rightarrow \infty$ may be taken in any meaningful or useful
way, but we will not pursue that here).

Since each state is a linear combination of the
$ \psi_{nm}$, one may also derive
the following general formula,
\beq
\psi^{(K)}_{nm} = 2^{-K/2} \sum_{j=0}^{2^K - 1} \ c^K_j \ \psi_{n, 2^K m + j }
\eeq
where
\beq
c^K_j = (-1)^j   \ \ \textrm{if \ \ $ 0 \le j \le 2^{K-1} - 1 $}
\eeq
and
\beq
c^K_j = -(-1)^j  \ \ \textrm{if $ 2^{K-1} \le j \le 2^K - 1 $}
\eeq
%
We also have
\beq
\chi^{(K)}_{nm} = 2^{-K/2} \sum_{j=0}^{2^K - 1} \ (-1)^j \ \psi_{n, 2^K m + j }
\eeq
This may also be written,
\beq
\chi^{(K)}_{nm} = 2^{-K/2}  \frac{ ( 1 - e^{2^{K+1} \pi i x / a} )} { ( 1 + e^{2 \pi i x/a} ) }
\psi_{n, 2^{K+1} m }
\label{2.12}
\eeq

The states $ | \psi^{(K)}_{nm} \rangle$ are not all obtained from a
single fiducial state, since we have
\beq
| \psi^{(K)}_{nm} \rangle = U^{(K)}_{nm} | \psi^K \rangle
\label{2.13}
\eeq
where
\beq
U^{(K)}_{nm} = U_{n, 2^K m }
\label{2.14}
\eeq
and
\beq
| \psi^K \rangle = | \psi^K_{00} \rangle
\eeq
There are therefore $N$ fiducial states for the set $\psi^{(K)}_{nm}$, $K=1, \cdots N$.

At some length, one can compute the averages and dispersions of $p$ and $x$ in the
fiducial states $ | \psi^K \rangle$. One obtains
\bea
\langle  p \rangle_K &=& \frac {2 \pi \hbar} {a} \left(
2^{K-1} - \half \right)
\label{2.16}\\
\langle x \rangle_K &=& 0
\label{2.17}\\
(\Delta p)^2_K &=& \left( \frac {2 \pi \hbar} {a} \right)^2 \frac{ (2^{2K} - 1)} {12}
\label{2.18} \\
(\Delta x)^2_K &=& \frac{ a^2} {12}
\label{2.19}
\eea
(Note that $ \langle p \rangle_K $ is not zero. Because there is more than
one fiducial state, it does not appear to be possible to shift the states so that
$  \langle p \rangle_K  = 0 $ in the fiducial states without spoiling orthogonality.)

The construction described above is concisely summarized as follows: For each
set of $2^N$ lattice points in the momentum direction, there are $2^N - 1 $
states with finite dispersion and just $1$ state (the remainder state), with
infinite dispersion.

Consider the following simple example to illustrate the construction.
Consider the $8$ lattice
points $m=0,1 \cdots 7$ in the momentum direction, so $N=3$. Then the $8$ states which
depend only on these points are
\bea
\psi^{(1)}_{n,0} &=&  \frac {1} {\sqrt{2}} \left( \psi_{n,0} + \psi_{n,1} \right) \nonumber \\
\psi^{(1)}_{n,1} &=&  \frac {1} {\sqrt{2}} \left( \psi_{n,2} + \psi_{n,3} \right) \nonumber \\
\psi^{(1)}_{n,2} &=&  \frac {1} {\sqrt{2}} \left( \psi_{n,4} + \psi_{n,5} \right) \nonumber \\
\psi^{(1)}_{n,3} &=&  \frac {1} {\sqrt{2}} \left( \psi_{n,6} + \psi_{n,7} \right)
\eea
\bea
\psi^{(2)}_{n,0} &=&  \frac {1} {2} \left( \psi_{n,0} - \psi_{n,1} - \psi_{n,2} + \psi_{n,3} \right) \nonumber \\
\psi^{(2)}_{n,1} &=&  \frac {1} {2} \left( \psi_{n,4} - \psi_{n,5} - \psi_{n,6} + \psi_{n,7} \right)
\eea
\beq
\psi^{(3)}_{n,0} =  \frac {1} {2^{3/2}} \left( \psi_{n,0} - \psi_{n,1} + \psi_{n,2} - \psi_{n,3}
- \psi_{n,4} +  \psi_{n,5} - \psi_{n,6} + \psi_{n,7} \right)
\eeq
\beq
\chi^{(3)}_{n,0} =  \frac {1} {2^{3/2}} \left( \psi_{n,0} - \psi_{n,1} + \psi_{n,2} - \psi_{n,3}
+ \psi_{n,4} -  \psi_{n,5} + \psi_{n,6} - \psi_{n,7} \right)
\eeq
There are $7$ finite dispersion states and one infinite dispersion state. It is easy to
see that they are orthonormal and form a complete set on these $8$ lattice points.
Note also that widths of the states $\psi^{(K)}_{nm}$ increases with $K$, but this
is in some sense offset by the fact that the states become progressively sparser.

So far we have been working with a complete set of states. We now come to the specific form of the
proposal to relax the requirement of using a complete set of states (in the construction
of commuting operators $\hat P$ and $\hat X$): we quite simply drop
the remainder states $ \chi^{(N)}_{nm} $, which have infinite dispersion, and use
only the incomplete set $ \psi^{(K)}_{nm}$, $K=1,2 \cdots N$, which have finite
dispersion. That is, in each set of $2^N$ lattice points in the momentum direction,
we use $2^N - 1 $ out of the $2^N$ states.

It seems likely that this approximation will be valid for sufficiently large $N$, for suitable
density matrices. We will show in the next section that reasonable results can be obtained
using the $ \psi^{(K)}_{nm}$ alone.  Here, we briefly look at the remainder
states $\chi^{(N)}_{nm}$ to see under what conditions they may be dropped.
Note first that the states $ \psi^{(K)}_{nm} (x)$ have the property that
they vanish at the end-points of the intervals, $x = na \pm a/2$
(as they must, so that their derivative does not have a $\delta$-function).
Not surprisingly therefore, the remainder states  $\chi^{(N)}_{nm}(x) $
become narrower and progressively more concentrated about the end-points
as $N \rightarrow \infty $, as one can see from Eq.(\ref{2.12}) (since they are in some
sense making up for the fact that the $ \psi^{(K)}_{nm} (x)$ vanish at
the end-points, and the whole set of states is complete). This suggests that,
under suitably coarse grained conditions, the behaviour of theses states
at single points will become insignificant. We will see this more explicitly
below.


We may now use the almost complete set of states to construct the
desired projection operators $E_{nm}$ localized on phase space cells.
We choose the phase space cells to have $2^N$ lattice points in
the $p$-direction and $1$ lattice point in the $x$-direction.
We have found $2^N -1 $ states with finite dispersion in each of
those cells. We therefore take the projector $E$ to be
\beq
E = \sum_{K=1}^N \sum_{m=0}^{2^{N-K} - 1} | \psi_{0m}^{(K)} \rangle \langle
\psi_{0m}^{(K)} |
\label{2.22}
\eeq
It is a sum over all states depending only on the lattice points
$0$ to $ 2^N - 1 $ (in the momentum direction).
Importantly, this is an {\it exact} projection operator which localizes
onto a region of phase space -- it satisfies
\beq
E^2 = E
\label{2.23}
\eeq
exactly. The projection operator for any other cell
is easily obtained by
unitary displacement, using steps of size $2^N$ lattice points in the
momentum direction (and single steps in the $x$-direction):
\beq
E_{nm} = U^{(N)}_{nm} E  \left(U^{(N)}_{nm} \right)^{\dag}
\label{2.24}
\eeq
These projectors clearly satisfy the exclusivity condition, Eq.(\ref{1.12}),
exactly, but do not satisfy the exhaustivity condition Eq.(\ref{1.13})
since we have
\beq
\sum_{nm} E_{nm} + \sum_{n,m}\ |\chi^{(N)}_{nm}\rangle \langle
{\chi^{(N)}_{nm}} | = 1
\label{2.25}
\eeq
But, as we have argued, we expect the $\chi $ terms to be negligible
under suitable conditions so
the projectors $E_{nm}$ should be almost exhaustive
\beq
\sum_{nm} E_{nm} \approx 1
\label{2.26}
\eeq
The approximate nature of this property
may not in fact matter for many practical applications. The phase
space projector onto a large cell $\Gamma$ in phase space is defined
by
\beq
E_{\Gamma} = \sum_{n,m \in \Gamma} E_{nm}
\eeq
The projector onto the region outside $\Gamma$ is then defined
to be
\beq
\bar E_{\Gamma} = 1 - E_{\Gamma}
\eeq
so they are trivially exhaustive, $ E_{\Gamma} + \bar E_{\Gamma} =1$.
The key point here is that the remainder
states dropped in the construction of $E$ have infinite dispersion so they
do not naturally belong in the construction of a phase space projector
for a finite region of phase space.

We will need some further properties of $E$. We have
\beq
\Tr E = 2^N - 1
\eeq
which means it projects onto a phase space
region of size $ (2^N -1 ) (2 \pi \hbar) $.
The object $ E / \Tr E $ may be thought of as a density operator
and we can compute averages and variances to see the properties of $E$.
We have
\bea
\langle \hat p \rangle_E &=& \frac{ \Tr (p E ) } { \Tr E}
\\
&=& \frac {2 \pi \hbar} {a} \left( 2^{N-1} - \half \right)
\eea
and
\bea
( \Delta p)^2_E & \approx & \sum_{K=1}^N \frac{ (\Delta p)^2_K } {2^K}
\\
& \approx & \frac {2^{N+1} \pi^2 \hbar^2 } { 3 a^2 }
\eea
where we show only the leading terms of large $N$.
We also have
\bea
\langle \hat x \rangle_E &=& 0
\\
(\Delta x)_E^2 & = & \frac {a^2 } {12}
\eea
Since $ \langle \hat p \rangle_E \ne 0 $, it is in fact useful to
perform a simple translation in momentum and
define a related projector $E'$ with all the same properties
as $E$ except that $ \Tr ( p E') = 0$. We will use this in what
follows.

The construction of an exact projector with the above properties is the main
achievement of this section and will be used to construct the commuting
$\hat X$ and $\hat P$ operators below.

Finally, we make two minor remarks.
First, note that position and momentum enter in the construction in very different
ways. However, the constant $a$ is arbitrary, so may be tuned to make the width
of the projector $E$ arbitrarily small in either the $x$ or $p$ direction.
Also, the states Eq.(\ref{2.1}) and (\ref{2.2}) are a Fourier transform pair,
so we could easily interchange them and start with a set of states that
have perfect localization in $p$, instead of $x$. It would of course be of interest
to find a construction in which $x$ and $p$ entered on an equal footing.

Second,
we note for comparison that Omn\`es has made extensive use of
phase quasi-projectors of the form
\beq
P_{\Gamma} = \int_{\Gamma} dp dq \ | p,q \rangle \langle p, q |
\eeq
where $ | p,q \rangle $ are phase space localized states, such as the coherent
states \cite{Omn2}. These have proved very useful for discussing emergent
classicality in quantum theory. However, in contrast to the projectors
constructed here, these are not exact projectors,
since they obey the approximate relation
\beq
P_{\Gamma}^2 \approx P_{\Gamma}
\eeq
(and so $P_{\Gamma}$ and $ 1 - P_{\Gamma}$ are only approximately
exclusive). It would be of interest to revisit some of Omn\`es results
using the exact projectors constructed here.

\section{Validity of Approximate Completeness}

Now we come to a crucial check of our approach, which is to determine the conditions
under which working with an approximately complete set of states gives reasonable
results.
The density matrix $\rho$ of the system satisfies $ \Tr \rho = 1$. If the states
$ | \psi^{(K)}_{nm} \rangle $ for $K = 1, \cdots N$ are approximately
complete, then we should have
\beq
\sum_{K=1}^N \sum_{nm} \ \langle  \psi^{(K)}_{nm} | \rho | \psi^{(K)}_{nm} \rangle
\approx 1
\label{3.1}
\eeq
which is the same as
\beq
\sum_{nm} \Tr \left( E_{nm} \rho \right) \approx  0
\eeq
We check this. It is most useful to exploit the Wigner representation,
Eqs.(\ref{X2.1}),(\ref{X2.2}) together with the property Eq.(\ref{X2.3}),
so we have
\beq
\sum_{K=1}^N \sum_{nm} \ \langle  \psi^{(K)}_{nm} | \rho | \psi^{(K)}_{nm} \rangle
=  2 \pi \hbar \sum_{K=1}^N \sum_{nm} \int dp dq  \ W^{(K)} (p,q)
\ W_{\rho} ( p + \frac{2^{K+1} \pi \hbar}{a} m, q+ n a )
\eeq
where $W_{\rho}$ is the Wigner function of $\rho$ and $W^{(K)}$ is the Wigner
function of $ | \psi^K \rangle $. Now the crucial step is to approximate the
discrete sum over $n,m$ with an integral over continuous variables,
$ \bar p = 2^{K+1} \pi \hbar m /a $, $ \bar q = n a $ and
we have
\bea
\sum_{nm} W_{\rho} ( p + \frac{2^{K+1} \pi \hbar}{a} m, q+ n a )
& \approx &\int \frac { d \bar p d \bar q} {2^{K+1} \pi \hbar} \  \ W_{\rho} ( p + \bar p, q + \bar q )
\label{3.5}
\\
& = & \frac{1} {2^{K+1} \pi \hbar}
\eea
This approximation is valid as long as the Wigner function of $\rho$
is slowly varying over phase space volumes of size $ 2^K ( 2 \pi \hbar)$.
Since $K$ runs up to $N$, we require slow variation on a scale of size
$2^N ( 2 \pi \hbar) $.
This is typically the case for density matrices that are sufficiently decohered,
as we saw in Section 2.
We now have
\bea
\sum_{K=1}^N \sum_{nm} \ \langle  \psi^{(K)}_{nm} | \rho | \psi^{(K)}_{nm} \rangle
& \approx & \sum_{K=1}^N \frac {1} {2^K}
\\
&=& 1 - \frac{1}{2^N}
\label{3.6}
\eea
Hence we do indeed get a result close to $1$ as long as $N$ is sufficiently large.

One other related result is worth recording here since it will be used in the next
section. Any density operator satisfies the relation
\beq
\int \frac {dk dq} {2 \pi \hbar} U^{\dag} (q,k) \rho U(q,k) = 1
\label{3.7}
\eeq
where $U(q,k)$ is the unitary shift operator in phase space. This is easily
proved using the Wigner representation above, since we have
\beq
\langle x | U^{\dag} (q,k) \rho U(q,k) | y \rangle
= \int dp \ e^{\ih p (x-y) } \ W_{\rho}  ( p - k, \frac {x+y} {2} - q )
\eeq
and integrating the right-hand side over $k$ and $q$ yields $2 \pi \hbar \delta (x-y)$.

One would expect that, for sufficiently slowly varying Wigner functions,
a discrete version of the result Eq.(\ref{3.7}) would hold. So suppose
instead of the operator $U(q,k)$, we use the operators
$ U^{(K)}_{nm}$ defined in Eq.(\ref{2.14}). We then have
\beq
\langle x | (U^{(K)}_{nm})^{\dag} \rho U^{(K)}_{nm} | y \rangle
= \int dp \ e^{\ih p (x-y) } \ W_{\rho}  ( p - \frac{2^{K+1} \pi \hbar}{a} m, \frac {x+y} {2} - na  )
\eeq
Using the same approximation and same steps as in Eq.(\ref{3.5})
to do the sum over $n,m$, it follows that
\beq
\sum_{nm} \ (U^{(K)}_{nm})^{\dag} \rho U^{(K)}_{nm}
\ \approx \ \frac {1} {2^K}
\label{3.10}
\eeq
and therefore
\beq
\sum_{K=1}^N \sum_{nm} \ (U^{(K)}_{nm})^{\dag} \rho U^{(K)}_{nm}
\ \approx \ 1 - \frac {1} {2^N}
\eeq

\section{Construction of Commuting Position and Momentum Operators}

We now use the projection operator $E'_{nm}$ to construct commuting
position and momentum operators, as outlined in the introduction.
They are,
\bea
\hat X &=& \sum_{nm} \ X_n \ E'_{nm}
\\
\hat P &=& \sum_{nm} \ P_m \  E'_{nm}
\eea
where $X_n = n a$ and $P_m = m \times 2^N  \left( 2 \pi \hbar/ a \right) $.
Clearly
\beq
[ \hat X, \hat P ] = 0
\eeq
as required.

We need to determine whether these operators are close to
the original canonical pair, $\hat x, \hat p$. To do this we need some measure
of distance, $ \parallel \hat P - \hat p \parallel $. We will define this by
\beq
\parallel \hP - \hp \parallel^2 = \Tr \left( ( \hP - \hp )^2 \rho \right)
\eeq
(and similarly for $\hat X$) where $\rho$ is the density operator of the
system, which we will assume is reasonably spread out in phase space.
A more general measure of distance would include some sort of optimization
over $\rho$, but, for the class of reasonably decohered density operators,
the result depends only weakly on $\rho$.

We have
\beq
\parallel \hP - \hp \parallel^2 = \Tr \left( (\hP^2 - 2 \hP \hp + \hp^2) \rho \right)
+ \Tr \left( \hP [ \hp, \rho] \right)
\label{6.3}
\eeq
where the operators $\hP$ are moved to the left in the first term
in preparation for inserting the explicit form for $\hP$ below
(since $[\hP, \hp] \ne 0 $).
It is useful to introduce
the notation ${\rm tr}( \cdots ) $ to denote a trace over the incomplete set
of states $ | \psi^{(K)}_{nm} \rangle$. We may then write
\beq
\parallel \hP - \hp \parallel^2 = {\rm tr} \left( (\hP^2 - 2 \hP \hp + \hp^2) \rho \right)
+ d_p^2
+ \Tr \left( \hP [ \hp, \rho] \right)
\label{6.4}
\eeq
where
\beq
d_p^2 = \Tr ( \hp^2 \rho) - {\rm tr} ( \hp^2 \rho )
\eeq
In the previous section we saw that $\Tr \rho$ and ${\rm tr} \rho $
are both very close (to order $2^{-N}$) and it is reasonable to
expect the same if $\rho$ is replaced with $\hp^2 \rho$, so $d_p^2$
will negligible and we drop it.

We also expect the term $ \Tr \left( \hP [ \hp, \rho] \right) $ to be small,
since a decohered $\rho$ will be approximately diagonal in momentum.
Inserting the explicit expression for $\hat P$, it is easily shown that
\beq
\Tr \left( \hP [ \hp, \rho] \right) = \sum_{nm} P_m {\rm Tr} \left(
[\hp, (U^{(N)}_{nm})^{\dag} \rho U^{(N)}_{nm} ] \right)
\label{6.8}
\eeq
Since we are working in the approximation in which $\rho$ is sufficiently
slowly varying that the sum over $n$ becomes an integral. It is then
easily seen (using Eq.(\ref{3.7}, for example) that, in this approximation,
the object
\beq
\sum_n (U^{(N)}_{nm})^{\dag} \rho U^{(N)}_{nm}
\eeq
is diagonal in momentum. Therefore the term Eq.(\ref{6.8}) is zero
to the approximation we are using.

Consider the one remaining term in Eq.(\ref{6.4}).
Inserting the explicit for for $\hP$, we have
\bea
\parallel \hP - \hp \parallel^2 &=& \sum_{nm} {\rm tr} \left( E'_{nm} (P_m - \hp)^2 \rho \right)
\\
&=& \sum_{nm} {\rm tr} \left( E' \hp^2 (U^{(N)}_{nm})^{\dag}\rho U^{(N)}_{nm}  \right)
\eea
From Eq.(\ref{3.10}), we have
\beq
\sum_{nm} ( U^{(N)}_{nm})^{\dag}\rho U^{(N)}_{nm} \approx \frac {1} {2^N}
\eeq
and so
\bea
\parallel \hP - \hp \parallel^2 & \approx & (\Delta p)_E^2
\\
& \approx & \frac {2^{N+1} \pi^2 \hbar^2 } { 3 a^2 }
\eea
where only the leading term for large $N$ is given.

A similar (and simpler) calculation shows that
\bea
\parallel \hat X - \hat x \parallel^2 & \approx & (\Delta x)^2_E
\\
&=& \frac{ a^2} {12}
\eea
Putting all these results together we obtain
\beq
\parallel \hat P - \hat p \parallel . \parallel \hat X - \hat x \parallel
\ \approx \ C \hbar
\label{4.10}
\eeq
where
\beq
C = 2^{N/2} \frac {\pi } { 3 \sqrt{2}}
\eeq
This result is valid in the approximation of large $N$ and for slowly varying density
operators.

Eqs. (\ref{3.6}) and (\ref{4.10}) show that $N$ needs to be chosen to be ``large'' to
get approximate completeness, yet ``small'' for the commuting operators
to be close to the original canonical pair. It seems likely, however,
that there is a range of intermediate values that will meet both
of these requirements. For example, take $N=20$. Then $C \sim 10^3$,
safely within the limit estimated in Eq.(\ref{1.3a}), so the difference
between the old and new operators will be completely invisible to
macroscropic observations. The error due to approximate completeness
is of order $ 2^{-N}$, which is about
$10^{-6}$. So there appears to be a large regime
in which the approach works well.

\section{Probabilities for Position and Momentum}

The results Eq.(\ref{3.6}) and Eq.(\ref{4.10}) are only indicative,
and the true test of closeness of the old and new operators is a
comparison of the probabilities for
$x,p$ and $X,P$.

Consider the probability that the variable $X$ lies in the range $\Delta_X$,
where $\Delta_X $ is the interval $[n_1 a, n_2 a]$, for some integers $n_1, n_2$.
The probability is
\beq
p(\Delta_X) = \Tr \left(  \rho E_{\Delta_X}   \right)
\eeq
where the projector $ E_{\Delta_X} $ is defined by
\beq
E_{\Delta_X} = \sum_{n \in \Delta_X} \sum_m E_{nm}'
\eeq
(and the loose notation $n \in \Delta_X $ means $ n \in [n_1, n_2]$).
The probability for lying outside the region $\Delta_X$
is defined using the projector $ 1 - E_{\Delta_X}$, so
that the probabilities add to $1$ exactly, and the approximate
completeness discussed earlier poses no problems. The probability
for $P$ is similarly defined, in terms of a projector $E_{\Delta_P}$
defined by
\beq
E_{\Delta_P} = \sum_{m \in \Delta_P} \sum_n E_{nm}'
\eeq

Both of these probabilities therefore involve the probability
$p_{nm}$ associated with our basic phase space cell of size $2^N (2 \pi \hbar)$,
given by
\beq
p_{nm} = \Tr ( E'_{nm}  \rho) = \Tr \left( E' \ (U^{(N)}_{nm})^{\dag}\rho U^{(N)}_{nm} \right)
\eeq
It is more usefully written in the Wigner representation,
\beq
p_{mn} = 2 \pi \hbar \int dp dq \ W_{E'} (p,q)\  W_{\rho} (p+ P_m ,q+ X_n)
\eeq
where, recall, $X_n = na $ and $P_m = m \times 2^N (2 \pi \hbar / a )$.

The probability for the variable $X$ is
\beq
p (\Delta_X)  = \sum_{n \in \Delta_X} \sum_m p_{nm}
\eeq
Since the Wigner function
of $\rho$ is assumed slowly varying, the sum over $m$ may
be approximated by an integral over $P_m$ regarded as a continuous
variable. The $p$ in $p+ P_m$ is absorbed into the integration and
we obtain,
\beq
p (\Delta_X)  \approx \frac{a} {2^N}  \sum_{n \in \Delta_X} \int dp dq \ W_{E'} (p,q) \ \rho ( q + X_n, q+ X_n )
\eeq
Now the $p$ integral may be carried out with the result
\beq
p (\Delta_X) \approx a  \sum_{n \in \Delta_X} \int dq \ \frac {\langle q | E' | q \rangle} {2^N}
\ \rho ( q + X_n, q+ X_n )
\eeq
Since $E'$ is phase space localized, one can see that
the first part of the integrand
is a smearing function peaked about $q=0$ and with width
$ ( \Delta q)^2 \approx a^2 / 12 $. It is normalized
to $1$ (for large $N$) when integrated over $q$ since $ \Tr E' = 2^N - 1$.
If we assume that
\beq
\Delta_X^2 \gg \frac{ a^2} {12}
\eeq
then the presence of the smearing function makes no difference and
we obtain
\beq
p (\Delta_X) \approx a  \sum_{n \in \Delta_X}
\ \rho ( na , na )
\eeq
If, as we assume, the density operator is sufficiently slowly varying
for the discrete sum to become an integral, we obtain
\beq
p( \Delta_X) \ \approx \ \int_{\Delta_X} dX \ \rho (X, X)
\eeq
the usual probability for position.

Similarly for $P$, with analogous approximations, we obtain
\bea
p( \Delta_P)   &=& 2^N \left( \frac {2 \pi \hbar} {a} \right)
\sum_{m \in \Delta_P}
\int dp \ \frac {\langle p | E' | p \rangle} {2^N}
\ \tilde \rho ( p + P_m, p+ P_m )
\cr
& \approx &
 2^N \left( \frac {2 \pi \hbar} {a} \right)
\sum_{m \in \Delta_P}
\ \tilde \rho (  P_m, P_m )
\eea
where $\tilde \rho (p,p')$ is the Fourier transform of the density
matrix $\rho (x,y)$.
Again, for slowly varying density operators the discrete sum becomes an integral
and we obtain
\beq
p (\Delta_P) \ \approx \ \int_{\Delta_P} d P \ \tilde \rho (P,P )
\eeq
which coincides with the usual probability for $p$.

We therefore find that the probabilities for $X,P$ coincide with those
for $x,p$ as long as the following conditions hold:

\begin{itemize}

\item[(i)] $N$ is sufficiently large that the errors $1/2^N$ are tolerably
small

\item[(ii)] The density operator $\rho$ is slowly varying on scales of
size $2^N (2 \pi \hbar)$.

\item[(iii)] The widths of the projections satisfy $\Delta_X \Delta_P \gg 2^N (2 \pi \hbar) $.

\end{itemize}

The key restriction is (ii), the restriction on the density operator. Eqs.(\ref{X2.9}) and (\ref{X2.10})
indicate that the density operator can easily become sufficiently broad for (ii) to be
satisfied. For example, for a time-evolving state, (ii) will be satisfied for
\beq
t \gg \left( \frac {\hbar} { \gamma k T} \right)^{1/2} \ 2^{N/2}
\eeq
This can easily be extremely short, even for large $N$. Similarly, for a state
close to thermal equilibrium, (ii) is satisfied for
\beq
\frac {k T} {\hbar \omega} \gg 2^{N}
\eeq
which is again easily satisfied.

\section{Summary and Discussion}

We have constructed a pair of commuting operators $\hat X, \hat P$ which, at sufficiently coarse grained
scales, are close (in a variety of ways)
to the canonical position and momentum operators $\hat x, \hat p$. These commuting operators offer a new
way of defining the relationship between approximate and exact decoherence.

There are two ways in which this programme could be developed and improved.
First of all, this paper has concentrated on the technicalities of constructing
$\hat X$ and $\hat P$. It would be useful to develop more details of the conceptual
framework in which they are used to discuss emergent classicality.

Secondly, the present approach works at the coarse grained scales of about
$ 2^N$ phase space cells. Although arguably small to classical eyes, there are
some ways in which this scale is quite large, and there are indications that
some version of the present approach should work at finer scales. For example, it
is known that the Wigner function (and also the $P$-function) become positive when coarse
grained over just one or two phase space cells (rather than $2^N$ cells, as here),
a feature that is often taken as indicator of approximate decoherence and emergent
classicality \cite{DiKi}. This suggests that there might be another way to construct commuting position
and momentum operators which does not require such a large amount of phase space coarse graining.
It would, for example, be of interest to see if von Neumann's original suggestion
(involving an explicit orthogonalization of the coherent states) can actually be made
to work.

These and related questions will be pursued in future publications.

\section*{Acknowledgements}

I am particularly grateful to Joshua Zak for many useful conversations over a long
period of time. I would also like to thank Jeremy Butterfield and John Cardy
for many useful comments, and Keith Hannabus for bringing Ref.\cite{HaRo}
to my attention. I would particularly like to thank Thomas Elze for putting together
such an interesting meeting.


\end{document}